\documentclass[12pt,draft,reqno,tbtags]{amsproc}
\usepackage{faaeng}

\newtheorem{theo}{Theorem}

\input{tcilatex}

\begin{document}
\author{M. V. Pavlov and S. P. Tsarev}
\title{Tri-Hamiltonian Structures of The Egorov Systems\\
[2pt] of Hydrodynamic Type\thanks{%
\email{maxim.pavlov@mtu-net.ru; tsarev@edk.krasnoyarsk.su}}}

\section{Introduction}

\label{vved}

The study of general Hamiltonian structures for systems of hydrodynamic
type, i.e., systems of the form
\begin{equation}
u_{t}^{i}=\sum_{k=1}^{N}\upsilon_{k}^{i}({u})u_{x}^{k},\qquad i=1,\dots,N,
\label{ur1a}
\end{equation}
was initiated by Dubrovin and Novikov~\cite{DN-1983} and continued by Mokhov
and Ferapontov~\cite{MF-90}. In the present paper, we prove a theorem on the
existence of three Hamiltonian structures for a diagonalizable Hamiltonian
hydrodynamic type system~\eqref{ur1a} having two physical symmetries, with
respect to the Galilean transformations and to scalings, and possesses some
additional properties, namely, the metric of the Hamiltonian structure has
the Egorov property and the matrix $((r^{i}-r^{k})\beta_{ik})$ (see~Sec.~\ref%
{osn} below) is semisimple. There are rather many known physically
meaningful systems of hydrodynamic type which have this form, including
averaged equations on $N$-phase solutions of the Korteweg--de Vries equation
(the Whitham equations) and the nonlinear Schr\"odinger equation as well as
the dispersionless limits of the vector nonlinear Schr\"odinger equation and
the vector long-short resonance equation. For the Whitham equations, local
multi-Hamiltonian structures were described in \cite{AP-94,Aleks-95} by the
algebro-geometric method. In the present paper, we construct third
Hamiltonian structures for all above-mentioned systems by a common
differential-geometric method. It turns out that to find out whether there
exist second and third Hamiltonian structures and determine their type
(local, nonlocal with a constant curvature metric, or general nonlocal), it
suffices to know the homogeneity degrees of the annihilators of the first
Hamiltonian structure. These degrees are usually \textit{a priori} known. We
also prove a simple criterion for the Egorov property of a metric (Theorem~%
\ref{th-ego}). Theorems on tri-Hamiltonian structures are stated and proved
for the general case in~Secs.~\ref{Egor}--\ref{osn} and are then applied to
the above-mentioned examples in the subsequent sections.

The theory considered here can be applied not only to integrable
diagonalizable systems of hydrodynamic type but also to the theory of
Frobenius manifolds and conformal topological field theory for the solution
of the Witten--Dijkgraaf--Verlinde--Verlinde associativity equations~\cite%
{Dubr-96}.

Let us briefly state some results needed in the sequel (see~\cite{Ts-Izv}).

The original diagonalizable hydrodynamic type system~\eqref{ur1a} has
Riemann invariants $r^{i}=r^{i}(u)$, $i=1,\dots,N$, i.e., variables in which
the velocity matrix $(v_k^i)$ of system \eqref{ur1a} is diagonal (unless
otherwise stated, summation over repeated indices is not assumed):
\begin{equation}
r_{t}^{i}=\upsilon^{i}({r})r_{x}^{i},\qquad i=1,\dots,N.  \label{ur1b}
\end{equation}
In the following, we assume that $\upsilon^{i}\ne\upsilon^{k}$ for $i\ne k$
and that system \eqref{ur1b} \textit{does not split} in the sense that $\pa%
_{i}\upsilon^{k}\ne 0$, $i\ne k$. Proceeding from $\upsilon^{i}({r})$, we
find the Lam\'e coefficients $H_{i}(u)$ as some solutions of the
overdetermined system
\begin{equation}
\pa_{k}\ln H_{i}=\Gamma_{ik}^{i}=\frac{\pa_{k}\upsilon^{i}}{%
\upsilon^{k}-\upsilon^{i}}, \qquad i\ne k.  \label{ur2}
\end{equation}
Here $\pa_{k}\equiv\pa/\pa r^{k}$ and the $\Gamma_{ik}^{i}$ are the
Christoffel symbols of the Levi-Civita connection corresponding to the
metric entering the Hamiltonian operator $\wh{A}^{ij}$ of the system (see %
\eqref{ur5*} and \eqref{A} below). We also find the rotation coefficients
(see~\cite{Ts-Izv,Darboux})
\begin{equation}
\beta_{ik}=\frac{\pa_{i}H_{k}}{H_{i}},\qquad i\ne k.  \label{ur3}
\end{equation}
The compatibility conditions for systems \eqref{ur2} or \eqref{ur3} have the
form of the \textit{semi-Hamiltonian} property \cite{Ts-DAN,Ts-Izv}
\begin{equation}
\pa_{i}\beta_{jk}=\beta_{ji}\beta_{ik}\iff \pa_{j}\,\frac{\pa_{k}\upsilon^{i}%
}{\upsilon^{k}-\upsilon^{i}} =\pa_{k}\,\frac{\pa_{j}\upsilon^{i}}{%
\upsilon^{j}-\upsilon^{i}}, \qquad i\ne j\ne k.  \label{ur3*}
\end{equation}
Moreover, the diagonal coefficients of system \eqref{ur1b} can be
represented in the form $\upsilon^{i}({r})=\ov{H}_{i}/H_{i}$, where the $%
H_{i}$ are the Lam\'e coefficients and the $\ov{H}_{i}$ are some solutions
of the system
\begin{equation}
\pa_{i}\ov{H}_{k}=\beta_{ik}\ov{H}_{i},\qquad i\ne k.  \label{ur3**}
\end{equation}
A local Hamiltonian structure of Dubrovin--Novikov type exists for a
semi-Hamiltonian system \eqref{ur1b} if and only if
\begin{equation}
\pa_{i}\beta_{ik}+\pa_{k}\beta_{ki}+\sum_{m\ne
i,k}\beta_{mi}\beta_{mk}=0,\qquad i\ne k.  \label{ur4a}
\end{equation}
In this case, the hydrodynamic type system \eqref{ur1a} or \eqref{ur1b} can
be rewritten in the Hamiltonian form
\begin{equation}
u_{t}^{i}=\{u^{i},H\}=\sum_{j=1}^{N}\wh{A}^{ij}\,\frac{\delta H}{\delta u^{j}%
}=\sum_{k=1}^{N}(\nabla^{i}\nabla_{k}h)u_{x}^{k}  \label{ur5*}
\end{equation}
with Hamiltonian operator
\begin{equation}
\wh{A}^{ij}=g^{ij}(u)\,\frac{d}{dx}-\sum_{s,k}g^{is}(u)
\Gamma_{sk}^{j}(u)u_{x}^{k},  \label{A}
\end{equation}
where $g^{ij}=g^{ji}$ is a nondegenerate zero curvature metric and the
connection $\Gamma_{sk}^{j}$ is symmetric and compatible with the metric.
(Here $H=\int h(u)\,dx$ is a hydrodynamic type Hamiltonian.) In the Riemann
invariants, the metric is diagonal, i.e., $g_{ii}=H_{i}^{2}$ ($g^{ii}g_{ii}=1
$), and the vanishing of the Riemann tensor is equivalent to~\eqref{ur4a}.
Thus, with each local Hamiltonian structure of system~\eqref{ur1b}, a system
of orthogonal curvilinear coordinates $r^{i}$ in a flat $N$-dimensional
pseudo-Euclidean space is associated. Note that some of the functions $H_{i}$
and $\beta_{ik}$ are pure imaginary in the case of pseudo-Riemannian
diagonal metrics. (It is metrics of this type that occur in applications, as
is elucidated by Theorem~\ref{thk} below.)

Each coefficient $H_{i}$ is determined by~\eqref{ur2} to within a
multiplication by a function of one variable, $H_{i}=%
\mu_{i}^{-1/2}(r^{i})H_{i}^{(\op{fix})}$. In this case, condition %
\eqref{ur4a} acquires the form
\begin{equation}
\frac{1}{2}\,\mu_{i}^{\prime}\beta_{ik}^{(\op{fix})}+\frac{1}{2}%
\,\mu_{k}^{\prime}\beta_{ki}^{(\op{fix})}+\mu_{i}\pa_{i}\beta_{ik}^{(\op{fix}%
)}+\mu_{k} \pa_{k}\beta_{ki}^{(\op{fix})}+\sum_{m\ne i,k}\mu_{m}\beta_{mi}^{(%
\op{fix})}\beta_{mk}^{(\op{fix})}=0.  \label{ur4b}
\end{equation}
Since system \eq{10} is linear in $\mu_i$, we can readily see that any two
Hamiltonian structures~\eq{9} associated with a given diagonal hydrodynamic
type system with nondegenerate metric $g_{ij}$ are automatically compatible.
As was shown in \cite{FP-95}, system \eqref{ur4b} for $\mu_{i}(r^{i})$ can
have at most $N+1$ linearly independent solutions. In the general case, it
has no solutions at all for a given semi-Hamiltonian system.

The hydrodynamic type system~\eqref{ur1a} can also have the nonlocal
Hamiltonian structure~\eqref{ur5*} with a nonlocal Hamiltonian operator~\cite%
{Fer-FAN}
\begin{equation}
\wh{A}^{ij}=g^{ij}d/dx-\sum_{s,k}g^{is}\Gamma_{sk}^{j}u_{x}^{k} +\sum_{\a%
,\beta,n,m}\e_{\a\beta}w_{n}^{i,(\a)}u_{x}^{n} (d/dx)^{-1}w_{m}^{j,(\beta
)}u_{x}^{m}.  \label{nonloc}
\end{equation}
Here the $w_{n}^{i,(\a )}(u)$ are the matrix coefficients of the
hydrodynamic type flows
\begin{equation}
u_{t_{(\a)}}^{i}=\sum_{j=1}^Nw_{j}^{i,(\a )}({u})u_{x}^{j}
\end{equation}
commuting with~\eqref{ur1a}. In the Riemann invariants, the coefficients $%
w_{n}^{i,(\a)}$ are diagonal, namely, $w_{i}^{i,(\a)}=H_{i}^{(\a)}\!/H_{i}$,
where $H_{i}^{(\a)}$ is a set of solutions of system~\eqref{ur3**} such that
\begin{equation}
\pa_{i}\beta_{ik}+\pa_{k}\beta_{ki}+\sum_{m\ne i,k}\beta_{mi}\beta_{mk}
=\sum_{\a,\beta}\e_{\a\beta}H_{i}^{(\a)}H_{k}^{(\beta)},\qquad i\ne k,
\label{urQQ}
\end{equation}
$\e_{\a\beta}=\e_{\beta\a}=\op{const}$. Thus, the metric $g^{ij}(u)$ in the
above nonlocal Hamiltonian operator is not flat. If Eq.~\eqref{urQQ}
contains a single term $H_{i}H_{k}$ formed of the Lam\'e coefficients $H_{i}=%
\sqrt{g_{ii}}$ of the \textit{given metric}, we obtain a constant curvature
metric \cite{MF-90}.

\section{The Egorov Systems of Hydrodynamic Type}

\label{Egor}

As is shown in \cite{Ts-DAN,Ts-Izv}, a diagonal semi-Hamiltonian system %
\eqref{ur1b} has infinitely many commuting hydrodynamic type flows
\begin{equation}
r_{y}^{i}=w^{i}({r})r_{x}^{i},\qquad i=1,\dots,N,  \label{ur8}
\end{equation}
whose coefficients $w^{i}({r})$ can be found as solutions of the consistent
overdetermined system of first-order linear partial differential equations
\begin{equation}
\pa_{k}w^{i}=\Gamma_{ik}^{i}(w^{k}-w^{i}),\qquad i\ne k.  \label{ur9}
\end{equation}
System \eqref{ur1b} also has infinitely many hydrodynamic conservation laws $%
dp({r})/dt=dq({r})/dx$, where the conservation law densities $p({r})$ can be
found from the consistent system
\begin{equation}
\pa_{i}\pa_{k}p=\Gamma_{ik}^{i}\pa_{i}p+\Gamma_{ik}^{k}\pa_{k}p,\qquad i\ne
k.  \label{ur10}
\end{equation}
The substitution $w^{i}=\wt{H}_{i}/H_{i}$ reduces system~\eqref{ur9} to the
form~\eqref{ur3**}, and the substitution $\pa_{i}p=\psi_{i}H_{i}$ reduces~%
\eqref{ur10} to the system
\begin{equation}
\pa_{k}\psi_{i}=\beta_{ik}\psi_{k},\qquad i\ne k,  \label{ur11}
\end{equation}
which is the adjoint of~\eqref{ur3**}.

\begin{pf}{Definition 1}
An orthogonal coordinate system associated with the diagonal
metric $g_{ii}=H_{i}^{2}$ is called a \textit{The Egorov
coordinate system} if
\begin{equation}
\beta_{ik}=\beta_{ki}. \label{ur6a}
\end{equation}
\end{pf}

In this case, the metric is \textit{potential}, i.e., $g_{ii}=\pa_ia(r)$ for
some function $a(r)$.

The corresponding hydrodynamic type systems will also be called
The Egorov systems. Condition~\eqref{ur6a} is not invariant with
respect to the natural transformation $r^{i}\to\varphi^{i}(r^{i})$
of Riemann invariants or the transformation
$H_{i}\to\mu_{i}^{-1/2}(r^{i})H_{i}$. An invariant condition
for a nonsplitting system ($\beta_{ik}\ne 0$) with $N\ge3$ (see \cite%
{Egorov,Darboux}) is given by the relation
\begin{equation}
\beta_{ik}\beta_{kj}\beta_{ji}=\beta_{ij}\beta_{jk}\beta_{ki},\qquad i\ne
j\ne k.  \label{ur6b}
\end{equation}
For a Egorov system of hydrodynamic type, the relations $\pa_{i}p=H_{i}\ov{H}%
_{i}$ hold and systems \eqref{ur3**} and \eqref{ur11} coincide. Integrable
systems of hydrodynamic type have infinitely many independent conservation
laws $\pa_{t}p_{k}({u})=\pa_{x}q_{k}({u})$, $k=1,2,\dots$, and hence can be
rewritten in the conservative form
\begin{equation}
\pa_{t}a^{\a}=\pa_{x}q^{\a}({a}),\qquad\a =1,\dots,N,  \label{ur1c}
\end{equation}
where the $a^{\a}=a^{\a}({u})$ are conservation law densities and the $q^{\a%
}({a})$ are the corresponding flows.

Let us prove the following \textit{criterion for the Egorov property} of
system \cite{Pav-Whitham}:

\begin{theo}
\label{th-ego} A diagonalizable nonsplitting semi-Hamiltonian system %
\eqref{ur1a} has a pair of conservation laws of the form
\begin{equation}
a_{t}=b_{x},\qquad b_{t}=c_{x}  \label{ur7}
\end{equation}
if and only if it is a Egorov system. In this case, in the Riemann
invariants, the relations $\pa_{i}a=H_{i}^{2}$, $\pa_{i}b=H_{i}\ov{H}_{i}$,
and $\pa_{i}c=\ov{H}_{i}^{2}$ hold, where $\ov{H}_{i}=\upsilon^{i}({r})H_{i}$%
.
\end{theo}

\begin{pf}{Proof}
Condition \eqref{ur7} can be rewritten as
\begin{alignat}2
\pa_{i}a &=\psi_{i}H_{i},&\qquad\pa_{i}b&=\psi_{i}\ov{H}_{i}, \\
\pa_{i}b &=\ov{\psi }_{i}H_{i},&\qquad
\pa_{i}c&=\ov{\psi}_{i}\ov{H}_{i},\label{ur8*}
\end{alignat}
where the $\psi_{i}$ and $\ov{\psi_{i}}$ are solutions of
system~\eqref{ur11}, whence it follows that
$\ov{\psi}_{i}H_{i}=\psi_{i}\ov{H}_{i}$. Differentiating
$\ov{\psi}_{i}=\psi_{i}\ov{H}_{i}/H_{i}$ with respect to the Riemann
invariant $r^{k}$ ($i\ne k$) and dividing the result by
$\upsilon_{i}-\upsilon_{k}=\ov{H}_{i}/H_{i}-\ov{H}_{k}/H_{k}$, we obtain
\begin{equation}
\frac{\psi_{k}}{H_{k}}\,\beta_{ik}=\frac{\psi_{i}}{H_{i}}\,\beta_{ki}
\label{ur9*}
\end{equation}
from \eqref{ur11}, which implies \eqref{ur6b} provided that the number
$N$ of equations is greater than~$2$. Note that, by virtue of~\eqref{ur2}
and~\eqref{ur3}, the nonsplitting condition ($\pa_{i}\upsilon^{k}\ne 0$,
$i\ne k$) implies that $\beta_{ki}\ne 0$ for $i\ne k$ and by virtue
of~\eqref{ur9*}, it guarantees that $\psi_{i}\ne 0$ and $\pa_{i}a\ne 0$
for any $i$. To prove the Egorov property for the case in which
system~\eqref{ur1a} contains only two equations, we set
$q_{i}=\psi_{i}/H_{i}=\ov{\psi }_{i}/\ov{H}_{i}$. It follows from
\eqref{ur8*} that $\pa_{i}(q_{k}H_{k}^{2})=\pa_{k}(q_{i}H_{i}^{2})$ and
$\pa_{i}(q_{k}\ov{H}_{k}H_{k})=\pa_{k}(q_{i}\ov{H}_{i}H_{i})$, whence,
using \eqref{ur9*}, we arrive at the relations
$H_{i}^{2}\pa_{k}q_{i}=H_{k}^{2}\pa_{i}q_{k}$ and
$\ov{H}_{i}H_{i}\pa_{k}q_{i}=\ov{H}_{k}H_{k}\pa_{i}q_{k}$. Dividing one
of these expressions by the other (under the assumption that
$\pa_{k}q_{i}\ne 0$, $i\ne k$), we obtain $\upsilon_{i}=\upsilon_{k}$,
which contradicts the assumption that $\upsilon_{i}\ne \upsilon_{k}$ for
$i\ne k$. Consequently, $\pa_{k}q_{i}=0$. Since the Riemann invariants
are defined only to within the transformation $r^{i}\to
\varphi^{i}(r^{i})$, we can always make the functions
$q_{i}(r^{i})=\psi_{i}/H_{i}$ in \eqref{ur9*} identically equal to unity.
Thus, the metric is potential and we have $\pa_{i}a=H_{i}^{2}$,
$\pa_{i}b=H_{i}\ov{H}_{i}$, and $\pa_{i}c=\ov{H}_{i}^{2}$.

Conversely, if the metric is potential, i.e., $H_{i}^{2}=\pa_{i}a$, then
conditions \eqref{ur6a} hold. In this case, we have $a_{t}=\pa_{i}a\cdot
r_{t}^{i}=H_{i}^{2}r_{t}^{i}=H_{i}^{2}\upsilon
^{i}({r})r_{x}^{i}=H_{i}\ov{H}_{i}r_{x}^{i}=b_{x}$ and
$b_{t}=\pa_{i}b\cdot r_{t}^{i}=H_{i}\ov{H}_{i}r_{t}^{i}=H_{i}\ov{H}_{i}
\upsilon^{i}({r})r_{x}^{i}=\ov{H}_{i}^{2}r_{x}^{i}= c_{x}$.\qed
\end{pf}

Note that the condition that the metric is flat has not been used in the
proof of the theorem. Egorov himself studied potential metrics with the
additional zero curvature condition. In this paper, potential metrics of
arbitrary curvature will be referred to as \textit{Egorov} metrics. Since
all commuting flows have the same metric (see~\eqref{ur2} and~\eqref{ur9}),
we arrive at the following result.

\begin{tm}{Corollary 1}
An arbitrary commuting flow \eqref{ur8} of a Egorov semi-Hamiltonian
system of hydrodynamic type has a pair of conservation laws of the
form~\eqref{ur7} $\pa_{y}a=\pa_{x}h$, $\pa_{y}h=\pa_{x}g$.
\end{tm}

It is obvious that if the hydrodynamic type system~\eqref{ur1b} has the
local Hamiltonian structure~\eqref{ur5*}, then the relationship $%
w_{k}^{i}(u)=\nabla^{i}\nabla_{k}p$ between the solutions of Eqs.~%
\eqref{ur10} for conservation law densities and the solutions of Eqs.~%
\eqref{ur9} for commuting flows holds. Writing the operation of covariant
differentiation $\nabla^{i}$ in full and substituting the expressions $w^{i}(%
{r})=\ov{H}_{i}/H_{i}$ and $\pa_{i}p=\psi_{i}H_{i}$ (in Riemann invariants),
we obtain a relation between the solutions of problem~\eqref{ur3**} and the
adjoint problem~\eqref{ur11},
\begin{equation}
\ov{H}_{i}=\pa_{i}\psi_{i}+\sum_{m\ne i}\beta_{mi}\psi_{m}.  \label{ur13a}
\end{equation}
If a hydrodynamic type system has a second local Hamiltonian structure,
then, as was already noted, the diagonal coefficients $g_{(2)}^{ii}$ of its
metric can differ only in factors depending on the corresponding Riemann
invariants, $g_{(2)}^{ii}=\mu_{i}(r^{i})g_{(1)}^{ii}$. Let us make a
transformation $r^{i}\to\varphi^{i}(r^{i})$ of the Riemann invariants that
gives $\mu_{i}(r^{i})\to r^{i}$ (under the assumption that $%
\mu_i^{\prime}(r^i)\ne0$), i.e., $g_{(2)}^{ii}=r^{i}g_{(1)}^{ii}$. By
analogy with \eqref{ur13a}, for the connection generated by the second flat
metric $g_{(2)}^{ii}$ we obtain
\begin{equation}
\wt{H}_{i}=\frac{1}{2}\,\psi_{i}+r^{i}\pa_{i}\psi_{i}+\sum_{m\ne
i}r^{m}\beta_{mi}\psi_{m}.  \label{ur13c}
\end{equation}
If $g_{(1)}^{ii}$ is an Egorov metric, then formulas \eq{25} and \eq{26}
imply that
\begin{equation}
\ov{H}_{i}=\delta\psi_{i},\qquad\wt{H}_{i}=(\wh{R}+1/2)\psi_{i},
\label{ur14}
\end{equation}
where $\delta =\sum \pa_{k}$ is a translation operator and $\wh{R}=\sum r^{k}%
\pa_{k}$ is a scaling operator (see~\cite{PavTs}).

Let us summarize the preceding. For an appropriate choice of Riemann
invariants, if an Egorov hydrodynamic type system has one local Hamiltonian
structure, then $\delta\beta_{ik}=\nobreak0$; in other words, the rotation
coefficients of orthogonal coordinate systems depend only on the differences
of Riemann invariants, $\beta_{ik}=\beta_{ik}(r^{m}-r^{n})$. If the Egorov
hydrodynamic type system has also a second local Hamiltonian structure with
metric $g_{(2)}^{ii}=r^{i}g_{(1)}^{ii}$, then Eqs.~\eqref{ur4b} and~%
\eqref{ur3*} imply the equivalent homogeneity condition $\wh{R}%
\beta_{ik}=-\beta_{ik}$; thus, the rotation coefficients of orthogonal
coordinate systems are homogeneous functions of order $-1$. As is shown in~%
\cite{Ts-Izv}, the Egorov property of the metric for a diagonalizable
Hamiltonian system essentially expresses the fact that the system is
invariant with respect to the group of Galilean transformations $(x,t)\to
(x-V\cdot t,t)$. The homogeneity of rotation coefficients in some well-known
examples corresponds to the natural physical condition of invariance with
respect to the choice of units of measurement, which results in the
homogeneity of $\upsilon_{i}(r)$. Thus, the presence of two local
Hamiltonian structures for physical examples of hydrodynamic type systems is
a manifestation of the simplest fundamental invariance conditions. We have
already noted that the corresponding Hamiltonian structures prove to be
compatible automatically.

In the following, for brevity we say ``homogeneous rotation coefficients'',
implying that the degree of homogeneity is always $-1$, $\wh{R}%
\beta_{ik}=-\beta_{ik}$. The corresponding curvilinear orthogonal coordinate
systems and diagonal hydrodynamic type systems~\eq{2} will also be simply
said to be homogeneous.

\section{Annihilators of the First Local Hamiltonian Structure and the
Signature of its Metric for the Homogeneous Egorov Systems}

\textit{Annihilators} of a local Hamiltonian structure are defined as
conservation law densities $a^{\a}(u)$ determined by the vanishing condition
for commuting flows:
\begin{equation}
0=\nabla^{i}\nabla_{k}a^{\a},\qquad\a=1,\dots,N.  \label{ur12b}
\end{equation}
The overdetermined system~\eqref{ur12b} has an $N$-dimensional solution
space, which specifies flat coordinates for the metric $g_{ii}=H_{i}^{2}$ of
the Hamiltonian structure. Introducing $\psi_{i}^{(\a)}=\pa_{i}a^{\a}/H_{i}$
in the Riemann invariants, we rewrite Eq.~\eqref{ur12b} in the form
\begin{equation}
\pa_{k}\psi_{i}^{(\a)}=\beta_{ik}\psi_{k}^{(\a)},\quad 0=\pa_{i}\psi_{i}^{(\a%
)}+\sum_{m\ne i}\beta_{mi}\psi_{m}^{(\a)},\qquad \a=1,\dots,N.  \label{ur15}
\end{equation}
The following two theorems were proved in~\cite{Ts-diss} and later presented
in~\cite{Dubr-96}.

\begin{theo}
The homogeneous solutions $\psi_{i}^{(\a)}$ of system~\eqref{ur15} for an
Egorov orthogonal coordinate system whose rotation coefficients are
homogeneous and depend only on the difference of the Riemann invariants are
eigenvectors of the skew-symmetric matrix $\boldsymbol{B}$ with entries $%
B_{ij}=(r^{i}-r^{j})\beta_{ij}$, which has constant eigenvalues.
\end{theo}

\begin{pf}{Proof} We have
\begin{align*}
\wh{R}\psi_{i}^{(\a)}&=r^{i}\psi_{i,i}^{(\a)}
+\sum_{m\ne i} r^{m}\pa_{m}\psi_{i}^{(\a)}\\
&=\sum_{m\ne i}
r^{m}\beta_{mi}\psi_{m}^{(\a)}-r^{i}\sum_{m\ne i} \beta_{mi}\psi_{m}^{(\a)}
=\sum_{m\ne i} (r^{m}-r^{i})\beta_{mi}\psi_{m}^{(\a)}=c_{\a}\psi_{i}^{(\a)}.
\end{align*}
The constancy of the eigenvalues $c_{\a}({r})$ of the matrix $\boldsymbol{B}$
can be proved as follows: $\pa_{i}(\wh{R}\psi_{k}^{(\a)})
=\wh{R}(\pa_{i}\psi_{k}^{(\a)})+\pa_{i}\psi_{k}^{(\a)}=
\wh{R}(\beta_{ki}\psi_{i}^{(\a)})+\beta_{ki}\psi_{i}^{(\a)}
=\wh{R}(\beta_{ki})\psi_{i}^{(\a)}+\beta_{ki}\wh{R}(\psi_{i}^{(\a)})
+\beta_{ki}\psi_{i}^{(\a)}=-\beta_{ki}\psi_{i}^{(\a)}
+c_{\a}\psi_{i}^{(\a)}\beta_{ki}+\beta_{ki}\psi_{i}^{(\a)}
=c_{\a}\psi_{i}^{(\a)}\beta_{ki}$, i.e., $\pa_{i}(c_{\a}\psi_{k}^{(\a)})
=\pa_{i}(c_{\a})\psi_{k}^{(\a)}+c_{\a}\beta_{ki}\psi_{i}^{(\a)}
=c_{\a}\psi_{i}^{(\a)}\beta_{ki}$, whence $\pa_{i}(c_{\a})\equiv0$.\qed
\end{pf}

The expressions $\psi_{i}^{(\a)}$ will also be referred to as \textit{%
annihilators}.

Since the homogeneity degrees of any conservation law densities, including
annihilators, are real numbers in the case of homogeneous real hyperbolic
systems \eqref{ur1b} considered here, it follows that the eigenvalues of the
\textit{skew-symmetric} matrix $\boldsymbol{B}$ are also real. This can be
accounted for by the fact that for this class of hydrodynamic type systems
the metric is necessarily pseudo-Euclidean (see Theorem~\ref{thk} below),
i.e., some Lam\'e coefficients $H_{i}=\sqrt{g_{ii}}$ and rotation
coefficients $\beta_{ik}$ are pure imaginary. Thus, the matrix $%
\boldsymbol{B}$ has both real and pure imaginary entries, and one can no
longer claim that there is a complete set of eigenvectors. For example, if $%
N=3$, then for an Egorov homogeneous metric of signature~$1$ one can choose
coefficients $\beta_{ij}$ such that $(r^1-r^2)^2\beta_{12}^{2}+(r^1-r^3)^2%
\beta_{13}^{2}+(r^2-r^3)^2 \beta_{23}^{2}=0$. (It is easily seen
that this condition is compatible with Eqs.~\eqref{DDD} for
$\beta_{ik}$, given below.) In this situation, the matrix
$\boldsymbol{B}$ has one eigenvector and two root vectors with
eigenvalue~$0$. Note that the set of The Egorov orthogonal
coordinate systems of given dimension with homogeneous rotation
coefficients depending only on the differences of Riemann
invariants is a finite-parameter family, and the description of
such systems can be reduced to that of solutions of the consistent
Pfaff system
\begin{equation}
\pa_{i}\beta_{jk}=\beta_{ji}\beta_{ik},\quad \delta \beta_{ik}=0,\quad \wh{R}%
\beta_{ik}=-\beta_{ik},\qquad i\ne j\ne k,  \label{DDD}
\end{equation}
whose solution can be specified by setting an \textit{arbitrary} symmetric
matrix $(\beta_{ik})$ (where the $\be_{ii}$ are undefined) at the initial
point $(r_{(0)}^{i})$.

However, all physical examples of diagonalizable homogeneous
Egorov systems known to the authors, as well as The Egorov
orthogonal coordinate systems arising in conformal topological
field theory, for which the rotation coefficients are homogeneous
and depend only on the differences of Riemann invariants, result
in matrices $\boldsymbol{B}$ with \textit{complete sets of
eigenvectors}. Moreover, as a rule, the homogeneity degrees of
annihilators of the first local Poisson bracket are \textit{a
priori} known for physical systems. Hence we assume throughout the
following that the \textit{semisimplicity condition} holds,
namely, the matrix $\boldsymbol{B}$ has a complete set of
eigenvectors at some point $(r^i_{(0)})$. As was shown
in~\cite{Dubr-96}, the description of semisimple solutions of system~%
\eqref{DDD} can be reduced to the solution of the sixth Painlev\'e equation.

\begin{theo}
For a semisimple matrix $\boldsymbol{B}$, all annihilators $\psi_{i}^{(\a)}$
corresponding to a flat homogeneous Egorov coordinate system can be taken in
the form of homogeneous functions,
\begin{equation}
\boldsymbol{B}\vec\psi^{(\a)}=\wh{R}\vec\psi^{(\a)}=c_{\a}\vec\psi^{(\a)},
\qquad\a=1,\dots,N,  \label{ur16}
\end{equation}
where the $c_{\a}$ \rom(the homogeneity degrees of the annihilators\rom) are
simultaneously the eigenvalues of the matrix $\boldsymbol{B}$.
\end{theo}

\begin{pf}{Proof}
For a symmetric matrix ($\beta_{ik}$), i.e., for The Egorov
metrics, the second relation in \eqref{ur15} becomes
$\delta\psi_{i}^{(\a)}=\pa_{i}\psi_{i}^{(\a)}+\sum_{m\ne i}
\beta_{im}\psi_{m}^{(\a)}=\pa_{i}\psi_{i}^{(\a)}+\sum_{m\ne
i}\pa_{m}\psi_{i}^{(\a)}=0$. Thus, the (possibly nonhomogeneous)
functions $\psi_{i}^{(\a)}$ satisfy the system
\begin{equation} \delta \psi_{i}^{(\a)}=0, \qquad
\pa_{i}\psi_{k}^{(\a)}=\beta_{ik}\psi_{i}^{(\a)}, \quad i \ne k.
\label{ur17*}
\end{equation}
One can readily verify that this system is consistent and the initial
data for it are given by $N$ constants, namely, the values of the
function $\vec\psi^{(\a)}$ at some point ($r_{(0)}^{i}$). Reproducing the
computations made in the proof of the preceding theorem, we see that
$\pa_{i}(\wh{R}\ov{\psi }_{k}-c\ov{\psi}_{k})=
\beta_{ik}(\wh{R}\ov{\psi}_{i}-c\ov{\psi }_{i})$ and
$\delta(\wh{R}\ov{\psi}_{k}-c\ov{\psi}_{k})=0$ for each solution
$\ov{\psi}_{i}$ and each $c=\op{const}$. Thus,
$\wt{\psi}_{i}=\wh{R}\ov{\psi}_{i}-c\ov{\psi}_{i}$ also satisfies
system~\eqref{ur17*}. Consequently, equipping system~\eqref{ur17*} with
the initial data $\vec\psi^{(\a)}(r_{(0)}^{i})$ forming a basis of
eigenvectors of the matrix $\boldsymbol{B}$ at this point
($r_{(0)}^{i}$), we see that the $\wt{\psi}_{i}^{(\a)}
=\wh{R}\ov{\psi}_{i}^{(\a)}-c_{\a}\ov{\psi}_{i}^{(\a)}$ are also
solutions of system~\eqref{ur17*} that vanish at the point
($r_{(0)}^{i}$). Thus, $\wt{\psi}_{i}^{(\a)}\equiv0$, as desired.\qed
\end{pf}

\begin{pf}{Definition 2}
The \textit{momentum density} $P$ for a system of hydrodynamic type with
local Hamiltonian structure is understood as the quadratic expression
\begin{equation}
P=\frac{1}{2}\sum_{\a,\be}g_{\a\beta}a^{\a}a^{\beta}, \label{ur17}
\end{equation}
where the nonsingular matrix $g_{\a\beta}$ is the (constant) metric of
system \eq{2} in flat annihilator coordinates~$a^{\a}$.
\end{pf}

It is obvious (see~\cite{DN-1983}) that in the annihilator coordinates $a^{\a%
}$ the hydrodynamic type system \eqref{ur1a} has the form
\begin{equation}
a_{t}^{\a}=\frac{d}{dx}\bigg(\sum_{\be=1}^Ng^{\a\beta}\frac{\pa h}{\pa %
a^{\beta}}\bigg),  \label{ur18}
\end{equation}
where $h(u)$ is the Hamiltonian density. The momentum density gives an
integral of an arbitrary hydrodynamic type Hamiltonian system \eqref{ur18}
with given metric $g^{\a\beta}$ and generates the trivial commuting flow $%
u^i_t=u^i_x$.

Note the following assertion (proved independently in several papers by
different authors).

\begin{theo}
If a \rom(not necessarily diagonalizable\rom) hydrodynamic type system %
\eqref{ur1a} has $N+1$ linearly independent hydrodynamic conservation laws
and one of their densities can be expressed quadratically via the others \rom%
(see~\eqref{ur17}\rom), then the system has the local Hamiltonian structure %
\eqref{ur18} with constant metric $g_{\a\beta}$.
\end{theo}

\begin{pf}{Proof}
Let $\pa_ta^{\a}=\pa_x q^{\a}$ and
$\pa_tP=\pa_{t}[\frac{1}{2}\sum_{\a,\be}g_{\a\beta}a^{\a}a^{\beta}]=\pa_{x}Q({a})$.
It follows that
\begin{equation*}
 dQ\,{=} \sum_{\a,\beta}g_{\a\beta}a^{\a}dq^{\beta}({a})
\,{=}\sum_{\a,\beta}d[g_{\a\beta}a^{\a}q^{\beta}]
-\sum_{\a,\be}g_{\a\beta}q^{\beta}da^{\a}.
\end{equation*}
Consequently, $\sum_\be g_{\a\beta}q^{\beta}\equiv \pa h/\pa a^{\a}$ for
some function $h(u)$, and hence system~\eqref{ur1c} acquires the
form~\eqref{ur18}.\qed
\end{pf}

The following theorem, generalizing Theorem~4 to the case of a constant
curvature metric, was stated in~\cite{Pav-Ellips} and proved in~\cite{Pav-02}%
.

\begin{theo}
Suppose that a hydrodynamic type system represented in the form of
conservation laws $\pa_{t}c^{\a}=\pa_{x}b^{\a}$ has an additional
conservation law density $p$ quadratically related with the densities $c^{\a}
$\rom:
\begin{equation}  \label{kvadr}
p-\frac{\e}{2}\,p^{2}=\frac{1}{2}\sum_{\a,\beta}\bar{g}_{\a\beta} c^{\a%
}c^{\beta},
\end{equation}
where $\bar{g}_{\a\beta}$ is a constant nonsingular matrix. Then in the
field variables $c^{\a}$ the system has the nonlocal Hamiltonian structure
\begin{equation}
c_{t}^{\a}=\pa_{x}\bigg[\sum_{\beta=1}^N(\bar{g}^{\a\beta}-\e c^{\a%
}c^{\beta}) \frac{\pa h}{\pa c^{\beta}}+\e c^{\a}h\bigg]  \label{curve}
\end{equation}
associated with the metric $g^{\a\beta}=\bar{g}^{\a\beta}-\e c^{\a}c^{\beta}$
of constant curvature $\e$, where $(\bar{g}^{\a\beta})$ is the inverse
matrix of $(\bar{g}_{\a\beta})$.
\end{theo}

\begin{pf}{Proof}
It follows from the formulas $\pa_{t}c^{\a}=\pa_{x}b^{\a}$ and
$p_{t}=\pa_{x}Q(c)$ and from \eqref{kvadr} that
$$
dQ=\sum_{\a,\beta}\bar{g}_{\a\beta}\frac{c^{\beta}}{1-\e p}\,db^{\a}.
$$
By setting $q^{\a}=c^{\a}/(1-\e p)$, we find that
$\sum_{\beta}\bar{g}_{\a\beta}b^{\beta}=\pa s/\pa q^{\a}$ with some
potential $s$. Hence we see that
$$
p_{t}=\pa_{x}\bigg[\sum_{\beta}q^{\beta}\frac{\pa s}{\pa
q^{\beta}}-s\bigg]\quad\text{and}\quad
c_{t}^{\a}=\pa_{x}\bigg[\sum_{\beta}\bar{g}^{\a\beta}\frac{\pa s}{\pa
q^{\beta}}\bigg].
$$
Since
$$
\frac{\pa s}{\pa q^{\a}}=(1-\e p)\bigg[\frac{\pa s}{\pa
c^{\a}}-\sum_{\beta,\gamma}\e c^{\gamma }\frac{\pa s}{\pa
c^{\gamma}}\,\bar{g}_{\a\beta}c^{\beta}\bigg],
$$
we readily obtain \eqref{curve} with $h=(1-\e p)s$.\qed
\end{pf}

Needless to say, the shift $p\to p+\op{const}$ makes relation~\eqref{kvadr}
purely quadratic. However, the form given in \eqref{kvadr} clarifies the
passage to the limit as $\e\to0$.

\begin{theo}
The following relations hold for the components of an arbitrary \rom(not
necessarily Egorov or homogeneous\rom) flat metric in flat annihilator
coordinates\rom:
\begin{equation}
\sum_{\a ,\beta } g_{\a\beta}\psi_{i}^{(\a)}\psi_{k}^{(\beta )}=\delta_{ik},
\qquad g^{\a\beta}=\sum^N_{i=1}\psi_{i}^{(\a)}\psi_{i}^{(\beta )}.
\label{ur19}
\end{equation}
\end{theo}

For the special case of an Egorov metric, this result was obtained in~\cite%
{19}.

\begin{pf}{Proof}
The momentum density $P$ (see~\eqref{ur17}) satisfies the relation
$$
\pa_{i}P=\Psi_{i}H_{i}=\sum_{\a,\beta}g_{\a\beta}a^{\beta}
\psi_{i}^{(\a)}H_{i}\implies\Psi_{i}=\sum_{\a,\beta}g_{\a\beta}
a^{\beta}\psi_{i}^{(\a)}.
$$
Differentiating it with respect to $r^{k}$ and using \eqref{ur11}, we
obtain
$\sum_{\a,\beta}g_{\a\beta}\psi_{i}^{(\a)}\psi_{k}^{(\beta)}\big|_{i\ne
k}=0$. Taking account of the fact that $P$ generates the flow
$u_{t}^{i}=u_{x}^{i}$ and differentiating with respect to $r^{i}$, we
obtain $\sum_{\a,\beta}g_{\a\beta}\psi_{i}^{(\a)}\psi_{i}^{(\beta)}=1$,
which precisely gives the first formula in~\eqref{ur19}. Multiplying it
by $\psi_{k}^{(\gamma)}$ and performing summation over $k$, we arrive at
the relation
$$
\sum_{k=1}^{N}\bigg(\sum_{\a,\beta}g_{\a\beta}\psi_{i}^{(\a)}\psi_{k}^{(\beta)}
\bigg)\psi_{k}^{(\gamma)}=\sum_{\a=1}^{N}\psi_{i}^{(\a)}
\bigg(\sum_{\beta=1}^{N}g_{\a\beta}\sum_{k=1}^{N}\psi_{k}^{(\beta)}
\psi_{k}^{(\gamma)}\bigg)=\psi_{i}^{(\gamma)}.
$$
Since the  vectors $\vec{\psi}^{(\a)}$ are linearly independent, it
follows that
$$
\sum_{\beta=1}^{N}g_{\a\beta}\bigg(\sum_{k=1}^{N}\psi_{k}^{(\beta)}
\psi_{k}^{(\gamma)}\bigg)=\delta_{\a}^{\gamma},
$$
i.e., we have arrived at the second formula in~\eqref{ur19}.\qed

Let us return to the case of an Egorov orthogonal coordinate system whose
rotation coefficients are homogeneous and depend only on the differences
of Riemann invariants. Then
\begin{align*}
\wh{R}g^{\a\beta}=0=\wh{R}\sum_{i=1}^{N}\psi_{i}^{(\a)}\psi_{i}^{(\beta)}
=\sum_{i=1}^{N}\psi_{i}^{(\a)}(\wh{R}\psi_{i}^{(\beta )})
+\sum_{i=1}^{N}\psi_{i}^{(\beta)}(\wh{R}\psi_{i}^{(\a)})
=(c_{\a}+c_{\beta})\sum_{i=1}^{N}\psi_{i}^{(\a)}\psi_{i}^{(\beta)}.
\end{align*}
It follows that either $c_{\a}+c_{\beta}\ne 0$, and then the
corresponding component of the matrix $(g^{\a\beta})$ is zero, or
$c_{\a}+c_{\beta}=0$, and then the corresponding component of the matrix
$(g^{\a\beta})$ is nonzero. (Some of the components can also be zero, but
there must be sufficiently many nonzero components to ensure that the
metric is nondegenerate.) Hence if the semisimple matrix $\boldsymbol{B}$
has at most one eigenvector with zero eigenvalue, then, after an
appropriate renumbering of the flat coordinates corresponding to the
homogeneous functions $\psi_{i}^{(\a)}$, the flat metric acquires the
block antidiagonal form
\begin{gather}
g^{\a\beta}=\begin{pmatrix}\text{\capt{-5pt}{3pt}{\huge0}} &
\begin{matrix}\ast&\ast\\
\ast&\ast\end{matrix}\\
\begin{matrix}
\ast&\ast\\
\ast&\ast
\end{matrix}&\text{\capt{-6pt}{3pt}{\huge0}}
\end{pmatrix},\qquad
g^{\a\beta}=\text{\Small$\begin{pmatrix} \text{\huge0} & \begin{matrix}
0\\[-5pt]
\vdots
\\[-2pt]
0
\end{matrix}
& \text{\huge$\ast$}\\
\begin{matrix}
0&\dots&0
\end{matrix}
&
\begin{matrix}
\pm 1
\end{matrix}
&
\begin{matrix}
0 & \dots & 0
\end{matrix}\\
\text{\huge$\ast$}&
\begin{matrix}
0 \\[-5pt]
\vdots \\[-2pt]
0\end{matrix} & \text{\huge0}
\end{pmatrix}$}.\label{g}\\
N=2n\hskip110pt N=2n+1\notag
\end{gather}
\end{pf}

Thus we have proved the following assertion.

\begin{theo}
\label{thk} If the multiplicity of the zero eigenvalue of the semisimple
matrix $\boldsymbol{B}$ does not exceed $1$ for a flat homogeneous Egorov
coordinate system, then the signature of the given metric $g^{\a\be}$ \rom%
(the difference between the numbers of positive and negative diagonal
coefficients in a diagonal form of the metric\rom) is minimal. \rom(It is
equal to $0$ for $N$ even and to $\pm 1$ for $N$ odd.\rom)
\end{theo}

\section{The third Hamiltonian structure of the Egorov homogeneous systems}

\label{osn}

As was noted in~Sec.~2, our physical examples of hydrodynamic type systems
have two local Hamiltonian structures owing to the homogeneity and the
invariance with respect to Galilean transformations.

It is known from linear algebra that a skew-symmetric matrix $\boldsymbol{B}$
has pairs of eigenvalues differing in sign. In what follows, we sometimes do
not distinguish between eigenvalues $c_{\a}$ of $\boldsymbol{B}$ differing
in sign.

Let us expand the square of $\boldsymbol{B}$ as follows:
\begin{align}
[\boldsymbol{B}^2]_{ik}&=\sum_{m\ne i,k}
(r^{m}-r^{i})(r^{m}-r^{k})\beta_{mi}\beta_{mk}  \notag \\
&=\bigg[\sum_{m\ne i,k}(r^{m})^{2}\beta_{mi}\beta_{mk}+(r^{i})^{2}\pa_{i}
\beta_{ik}+(r^{k})^{2}\pa_{k}\beta_{ki}+r^{i}\beta_{ik} +r^{k}\beta_{ki}%
\bigg]  \notag \\
&\qquad-(r^{i}+r^{k})\bigg[\sum_{m\ne i,k}r^{m}\beta_{mi}\beta_{mk}+r^{i} \pa%
_{i}\beta_{ik}+r^{k}\pa_{k}\beta_{ki}+\frac{1}{2}\,\beta_{ik} +\frac{1}{2}%
\,\beta_{ki}\bigg]  \notag \\
&\qquad+r^{i}r^{k}\bigg[\sum_{m\ne i,k}\beta_{mi}\beta_{mk}+\pa_{i}
\beta_{ik}+\pa_{k}\beta_{ki}\bigg] +\frac{1}{2}(r^{i}-r^{k})(\beta_{ki}-%
\beta_{ik}).  \label{B-2}
\end{align}

\begin{theo}
\label{th-flat} For even $N=2n$, a homogeneous Egorov diagonal hydrodynamic
type system \eqref{ur1b} with flat metric $g_{(1)}^{ii}$ and second flat
metric $g_{(2)}^{ii}=r^ig_{(1)}^{ii}$ has a third local Hamiltonian
structure with metric $g_{(3)}^{ii}=(r^{i})^{2}g_{(1)}^{ii}$ if and only if
the matrix $\boldsymbol{B}$ is nonsingular and semisimple and has only one
pair of eigenvalues $\pm c$.
\end{theo}

\begin{pf}{Proof} In this case, $\boldsymbol{B}^{2}=c^{2}E$, where
$E$ is the identity matrix, since an application of $\boldsymbol{B}^{2}$
to any of its eigenvectors $\psi_i^{(\a)}$ gives $c^2\psi_i^{(\a)}$\!. On
the other hand, it follows from \eqref{B-2} that in the presence of two
local Hamiltonian structures determined by metrics $g_{(1)}^{ii}$ and
$g_{(2)}^{ii}=r^ig_{(1)}^{ii}$ (cf.~\eqref{ur4b} with $\mu_i=r^i$), one
has
$$
[\boldsymbol{B}^2]_{ik}= \sum_{m\ne
i,k}(r^{m})^{2}\beta_{mi}\beta_{mk}+(r^{i})^{2}\pa_{i}
\beta_{ik}+(r^{k})^{2}\pa_{k}\beta_{ki}+r^{i}\beta_{ik}+r^{k}
\beta_{ki}=0
$$
provided that $i\ne k$, which precisely implies the existence of a third
local Hamiltonian structure (cf.~\eqref{ur4b} with $\mu_i = (r^i)^2$).

Conversely, if condition~\eqref{ur4b} with $\mu_i = (r^i)^2$ holds, then
the matrix $\boldsymbol{B}^2$ is diagonal. Since system \eq{2} does not
split, it follows that $\beta_{ik} \ne 0$ for $i\ne k$. By~\eqref{ur11},
it is obvious that none of the components of any homogeneous annihilator
$\psi_i^{(\a)}$ vanishes identically. Since
$\boldsymbol{B}^2\psi_i^{(\a)} = c_{\a}^2\psi_i^{(\a)}$, we see that the
diagonal matrix $\boldsymbol{B}^2$ has identical diagonal entries
$c_{\a}^2$. One can readily prove that the matrix $\boldsymbol{B}$ is
semisimple by applying $\boldsymbol{B}^2$ to root vectors; namely, if
$\boldsymbol{B}\vec v=c_{\a}\vec v+\vec\psi^{(\a)}$, then
$\boldsymbol{B}^2\vec v = c^2_{\a}\vec v + 2c_{\a}\vec\psi^{(\a)} \ne
c^2_{\a}\vec v$.\qed
\end{pf}

\begin{theo}
\label{th-const} For odd $N=2n+1$, a homogeneous Egorov diagonal
hydrodynamic type system \eqref{ur1b} with flat metric $g_{(1)}^{ii}$ and
second flat metric $g_{(2)}^{ii}=r^ig_{(1)}^{ii}$ has a third nonlocal
Hamiltonian structure with constant curvature metric $%
g_{(3)}^{ii}=(r^{i})^{2}g_{(1)}^{ii}$ if and only if the matrix $%
\boldsymbol{B}$ is semisimple and has a simple eigenvalue $c_{(0)}=0$ and a
single pair of nonzero eigenvalues $c_{(\pm k)}=\pm c$ \rom($k=1,\dots,n$\rom%
) and the metric $g_{(1)}^{ii}$ itself is homogeneous of degree $0$ and
depends only on the differences of the Riemann invariants $r^{i}$.
\end{theo}

\begin{pf}{Proof} In this case,
$[\boldsymbol{B}^2]_{ij}=c^{2}\delta_{ij}-
(c^{2}\!/g^{00})\psi_{i}^{(0)}\psi_{j}^{(0)}$, where $\psi_{i}^{(0)}$ is
a basis element in the kernel of the matrix $\boldsymbol{B}$,
$\wh{R}\psi_{i}^{(0)}=\boldsymbol{B}\psi_{i}^{(0)}=0$, and
$g^{00}=\sum_{i}\psi_{i}^{(0)}\psi_{i}^{(0)}$ is the corresponding
element of the metric in flat coordinates. One can readily verify this
identity by applying $\boldsymbol{B}^2$ to $\psi_{i}^{(\a)}$ with regard
to the form~\eqref{g} of the metric in flat coordinates. Using the
expansion~\eqref{B-2} once more, we obtain
$$
\sum_{m\ne i,k}(r^{m})^{2}\beta_{mi}\beta_{mk}
+(r^{i})^{2}\pa_{i}\beta_{ik}+(r^{k})^{2}\pa_{k}\beta_{ki}
+r^{i}\beta_{ik}+r^{k}\beta_{ki}=4c^{2}H_{i}^{(0)}H_{k}^{(0)},
$$
since $\psi_{i}^{(0)}=2H_{i}^{(0)}$ (cf.~\eqref{ur14}). This just means
that there exists a third (this time, nonlocal) Hamiltonian structure
determined by the metric $g_{(3)}^{ii}=(r^{i})^{2}g_{(1)}^{ii}$ of
constant curvature $c$, since the expressions \eqref{urQQ} computed for
$g_{(3)}^{ii}$ read
$$
\frac{1}{r^{i}r^{k}}[\boldsymbol{B}^2]_{ik}
=\frac{4c^{2}}{r^{i}r^{k}}\,H_{i}^{(0)}H_{k}^{(0)},\qquad i\ne k.
$$
The existence of the corresponding Hamiltonian for this nonlocal
Hamiltonian structure readily follows from the results in~\cite{MF-90}.

Conversely, if an Egorov diagonal system with two flat metrics $g_{(1)}^{ii}$
and $g_{(2)}^{ii}=r^ig_{(1)}^{ii}$ and with homogeneous functions $\be_{ik}$
has a third nonlocal Hamiltonian structure with constant curvature metric
$g_{(3)}^{ii}=(r^{i})^{2}g_{(1)}^{ii}$, then
$[\boldsymbol{B}^2]_{ij}/(r^{i}r^{k})=d_{i}(r)
\delta_{ij}-c\wt{H}_{i}\wt{H}_{j}/(r^{i}r^{k})$, where
$\wt{H}_{i}(r)=1/\sqrt{\smash[b]{g_{(1)}^{ii}}}$ and the diagonal entries
$d_i(r)$ are some functions. Since the coefficients $\beta_{ik}$ are
homogeneous and invariant with respect to the shift $r^{i}\to
r^{i}+\op{const}$, it follows that all products $\wt{H}_{i}\wt{H}_{j}$ are
homogeneous of degree~$0$ and invariant with respect to the shift.
Consequently, expressing the functions $\wt{H}_{i}$ themselves via these
products, we find that the metric $g_{(1)}^{ii}$ is homogeneous of degree~$0$
and depends only on the differences of the Riemann invariants $r^{i}$. Applying
the matrix $\boldsymbol{B}^{2}$ to any eigenvector or root vector, we readily
find that all nonzero eigenvalues are the same and that there are no root
vectors.\qed
\end{pf}

\begin{theo}
A homogeneous Egorov diagonal hydrodynamic type system \eqref{ur1b} with
flat metric $g_{(1)}^{ii}$ and second flat metric $%
g_{(2)}^{ii}=r^ig_{(1)}^{ii}$ has a third nonlocal Hamiltonian operator of
the general form \eqref{nonloc} with metric $%
g_{(3)}^{ii}=(r^{i})^{2}g_{(1)}^{ii}$ provided that the matrix $%
\boldsymbol{B}$ is semisimple.
\end{theo}

\begin{pf}{Proof}
In this general case, one has
$$
[\boldsymbol{B}^2]_{ij}=c_{1}^{2}\delta_{ij}-
\sum_{s}c_{1}^{2}\psi_{i}^{(0,s)}\psi_{j}^{(0,s)} +\sum_{\a\ne
1}(c_{\a}^{2}-c_{1}^{2})(\psi_{i}^{({\a})}\psi_{j}^{({-\a})}
+\psi_{i}^{({-\a})}\psi_{j}^{({\a})}),
$$
where the annihilators $\psi_{i}^{(0,s)}$ are basis elements of the kernel of
the matrix $\boldsymbol{B}$, the $\psi_{i}^{({\a})}$ are its eigenvectors, and
the $\psi_{i}^{({-\a })}$ are its eigenvectors with opposite eigenvalues
arranged in a way such that the metric~\eqref{g} is diagonal for
$\psi_{i}^{(0,s)}$ and antidiagonal for $\psi_{i}^{({\a})},\psi_{i}^{({-\a})}$,
i.e., $\sum_{i} \psi_{i}^{({\a})}\psi_{i}^{({-\a})}=\nobreak1$; here we have
chosen one of the eigenvalues $c_{1}$ as the ``main'' diagonal entry of the
matrix $\boldsymbol{B}^{2}$. This relation, just as in the preceding cases, can
readily be verified by a straightforward substitution of the eigenvectors
of~$\boldsymbol{B}$. Thus, we can use~\eqref{ur14} and write out the
corresponding expansion \eqref{urQQ} guaranteeing the existence of the
Hamiltonian operator \eqref{nonloc}. If $\boldsymbol{B}$ has the eigenvalues
$\pm1/2$ (see~Secs.~5 and~6 below), one should take $c_1=-1/2$ to express the
desired annihilators $\psi_i^{(\a)}$ via the corresponding $H_i^{(\a)}$ with
the help of~\eq{27}.\qed
\end{pf}

Note that $H_{i}^{(0)}=\psi_{i}^{(0)}\!/2$ can be found explicitly without
quadratures even in the case of an arbitrary metric with homogeneous $%
\beta_{ik}$. For example, for $N=3$ we obtain $\psi_{i}^{(0)}=r^{j}%
\beta_{kj}-r^{k}\beta_{jk}$, and for $N=5$ one has $%
\psi_{i}^{(0)}=a_{jk}a_{lm}+a_{jl}a_{mk}+a_{jm}a_{kl}$, where $%
a_{jk}=r^{j}\beta_{kj}-r^{k}\beta_{jk}$. If the metric has the Egorov
property, then we obtain $\psi_{i}^{(0)}=(r^{j}-r^{k})\beta_{jk}$ for $N=3$
and $\psi_{i}^{(0)}=(r^{j}-r^{k})(r^{l}-r^{m})\beta_{jk}\beta_{lm}+
(r^{j}-r^{l})(r^{m}-r^{k})\beta_{jl}\beta_{mk}+(r^{j}-r^{m})(r^{k}-r^{l})
\beta_{jm}\beta_{kl}$ for $N=5$. In a similar way, one writes out $%
\psi_{i}^{(0)}$ for $N=2n-1>5$.

In the subsequent sections, we show how to establish the existence of second
and third Hamiltonian structures for hydrodynamic type systems that are not
written in Riemann invariants.

\section{Averaged $\boldsymbol{N}$-Phase Solutions of the Korteweg--de Vries
Equation\newline
and the Nonlinear Schr\"odinger Equation}

$N$-phase solutions of the KdV equation were averaged by the Whitham method
in \cite{FFM}. However, the Hamiltonian formalism for the averaged equations
was developed later in \cite{DN-1983,AP-94,Aleks-95}. The hydrodynamic type
systems obtained by averaging inherit Galilean invariance, Hamiltonian
property, and homogeneity. Moreover, since the KdV equation is Galilean
invariant and hence has a pair of local Hamiltonian structures and a pair of
conservation laws (see Theorem~1, Eq.~\eq{21})
\begin{equation*}
\pa_{t}u=\pa_{x}[u^{2}+\e^{2}u_{xx}],\qquad \pa_{t}[u^{2}+\e^{2}u_{xx}]=\pa%
_{x}[\tfrac{4}{3}u^{3}-3\e^{2}u_{x}^{2} +2\e^{2}(u^{2})_{xx}+\e^{4}u_{xxxx}],
\end{equation*}
it follows that after the averaging on an $N$-phase solution the resulting ($%
2N+1$)-component hydrodynamic type system is also Galilean invariant and has
an Egorov metric (by Theorem~\ref{th-ego}) that is homogeneous and depends
only on the differences of Riemann invariants. Hence, it has a pair of local
Hamiltonian structures. The annihilators of the first local Hamiltonian
structure of the averaged KdV equation comprise the averaged annihilator of
the first local Hamiltonian structure of the KdV equation itself as well as $%
N$ wave numbers (corresponding to the $N$ phases $\theta_{i}=k_{i}x-%
\omega_{i}t$ of the quasiperiodic solution) and $N$ corresponding partial
derivatives of the averaged Lagrangian with respect to these wave numbers. A
straightforward computation shows that the homogeneity degree of the
functions $\psi_{i}^{(0)}$ corresponding to the averaged annihilator of the
first local Hamiltonian structure is $0$, the homogeneity degree of the
functions $\psi_{i}^{(k)}$, $k=1,\dots,N$, corresponding to the wave numbers
is $-1/2$, and the homogeneity degree of the functions $\psi_{i}^{(N+k)}$, $%
k=1,\dots,N$, corresponding to the partial derivatives of the averaged
Lagrangian with respect to these wave numbers is~$1/2$. Thus (by Theorem~\ref%
{th-const}) the averaged KdV equations have a third nonlocal Hamiltonian
structure associated with a constant curvature metric (see~\cite{Pav-02}).
(The existence of the first two local Hamiltonian structures was established
earlier in \cite{DN-1983}.)

Multiphase solutions of the nonlinear Schr\"odinger equation were averaged
in the diploma paper of the first author. Later, this result was published
in~\cite{Pav-1987}. It was obtained independently in~\cite{FL}. It was
proved that averaged $N$-phase solutions of the NLS equation have three
local Hamiltonian structures and a fourth, nonlocal Hamiltonian structure
defined by a differential-geometric Poisson bracket with constant curvature
metric. The canonical form of the first three Hamiltonian structures in flat
coordinates was given in~\cite{Aleks-95} and~\cite{AP-94}. The nonlinear
Schr\"odinger equation written out in the Hasimoto form is Galilean
invariant and has three local Hamiltonian structures and the pair of
conservation laws
\begin{equation*}
\pa_{t}|u|^{2}=i\pa_{x}[u\bar{u}_{x}-\bar{u}u_{x}],\qquad i\pa_{t}[u\bar{u}%
_{x}-\bar{u}u_{x}]=\pa_{x}[|u|^{4}+4|u_{x}|^{2} -(|u|^{2})_{xx}].
\end{equation*}
Reproducing the above argument for the KdV equation in the case of the
nonlinear Schr\"odinger equation, one can also show that these properties
are inherited under $N$-phase averaging (see~\cite{Pav-Whitham}), which is a
good illustration of Theorems~\ref{th-ego} and~\ref{th-flat}.

\section{The Benney--Zakharov and Yajima--Oikawa--Mel\mz nikov Systems%
\newline
(the Dispersionless Limit)}

The Benney system (Zakharov's reduction; see~\cite{Ben,Zakharov})
\begin{equation}
u_{t}^{k}=\pa_{x}\bigg[\frac{(u^{k})^{2}}{2} +\sum_{m=1}^N\eta^{m}\bigg]%
,\qquad \eta_{t}^{k}=\pa_{x}(u^{k}\eta^{k})  \label{BZ}
\end{equation}
is Galilean invariant ($u^{k}\to u^{k}+c$, $x\to x-ct$) and has a pair of
conservation laws of the form (see Theorem~1, Eq.~\eq{21})
\begin{align*}
\pa_{t}\bigg(\sum_{m=1}^N\eta^{m}\bigg) &=\pa_{x}\bigg(\sum_{m=1}^Nu^{m}%
\eta^{m}\bigg), \\
\pa_{t}\bigg(\sum_{m=1}^Nu^{m}\eta^{m}\bigg) &=\pa_{x}\bigg[%
\sum_{m=1}^N(u^{m})^{2}\eta^{m} +\frac{1}{2}\bigg(\sum_{m=1}^N\eta^{m}\bigg)%
^{2}\bigg]
\end{align*}
and the first local Hamiltonian structure
\begin{equation*}
u_{t}^{k}=\pa_{x}\,\frac{\delta H}{\delta\eta^{k}},\qquad \eta_{t}^{k}=\pa%
_{x}\frac{\delta H}{\delta u^{k}}
\end{equation*}
determined by the following homogeneous Egorov metric depending on the
differences of Riemann invariants (in the Riemann invariants $\lambda^{k}$, $%
k=1,\dots,2N$; see~\cite{Ts-Izv}):
\begin{equation*}
g^{ii}=\sum_{m=1}^N\frac{\eta^{m}}{(\mu^{i}+u^{m})^{3}},
\end{equation*}
where the characteristic velocities $\mu^{i}$ and the Riemann invariants $%
\lambda^{i}$ can be found from the system (see~\cite{Gibbons})
\begin{equation*}
\lambda^{i}=\mu ^{i}+\sum_{m=1}^N\frac{\eta^{m}}{\mu^{i}+u^{m}},\qquad
1=\sum_{m=1}^N\frac{\eta^{m}}{(\mu^{i}+u^{m})^{2}}.
\end{equation*}
Comparing the homogeneity degrees of the flat coordinates $(u^{k},\eta^{k})$
and the metric $g^{ii}$, we see that the homogeneity degree of the functions
$\psi_{i}^{(k)}$, $k=1,\dots,N$, corresponding to the first half of flat
coordinates $u^{k}$ is $-1/2$ and the homogeneity degree of the functions $%
\psi_{i}^{(N+k)}$, $k=1,\dots,N$, corresponding to the second half of flat
coordinates $\eta^{k}$ is~$1/2$. It follows by Theorems~\ref{th-ego} and~\ref%
{th-flat} that the Benney--Zakharov solution~\eqref{BZ} has also second and
third local Hamiltonian structures (see~\cite{PavTs,Pav-Benney}).

In the dispersionless limit, the Yajima--Oikawa system (known as the
long-short resonance) generalized to the $N$-component case (Mel'nikov's
system~I; see~\cite{Oevel}) has the form
\begin{equation*}
u_{t}^{k}=\pa_{x}\bigg[\frac{(u^{k})^{2}}{2}+w\bigg],\quad \eta_{t}^{k}=\pa%
_{x}(u^{k}\eta^{k}),\quad w_{t}=\pa_{x}\bigg[\sum_{m=1}^N\eta^{m}\bigg].
\end{equation*}
It is also Galilean invariant ($w\to w+c$, $x\to x-ct$) and has a pair of
conservation laws of the form
\begin{equation*}
\pa_{t}w=\pa_{x}\bigg(\sum_{m=1}^N\eta^{m}\bigg),\qquad \pa_{t}\bigg(%
\sum_{m=1}^N\eta^{m}\bigg) =\pa_{x}\bigg[\sum_{m=1}^Nu^{m}\eta^{m}\bigg]
\end{equation*}
prescribed in Theorem~\ref{th-ego} and the first local Hamiltonian structure
\begin{equation*}
u_{t}^{k}=\pa_{x}\frac{\delta H}{\delta\eta^{k}},\qquad \eta_{t}^{k}=\pa%
_{x}\,\frac{\delta H}{\delta u^{k}},\qquad w_{t}=\pa_{x}\,\frac{\delta H}{%
\delta w}
\end{equation*}
determined by the following homogeneous Egorov metric depending on the
differences of Riemann invariants (in the Riemann invariants~$\lambda^{k}$, $%
k=1,,\dots,2N+1$):
\begin{equation*}
g^{ii}=1-2\sum_{m=1}^N\frac{\eta^{m}}{(\mu_{i}+u^{m})^{3}},
\end{equation*}
where the characteristic velocities $\mu_{i}$ and the Riemann invariants $%
\lambda_{i}$ can be found from the system
\begin{equation*}
\lambda^{i}=\frac{\mu_{i}^{2}}{2}+w-\sum_{m=1}^N\frac{\eta ^{m}}{%
\mu_{i}+u^{m}},\qquad \mu_{i}+\sum_{m=1}^N\frac{\eta^{m}}{(\mu_{i}+u^{m})^{2}%
}=0.
\end{equation*}
Comparing the homogeneity degrees of the flat coordinates $(w,u^{k},\eta^{k})
$ and the metric $g^{ii}$ we see that the homogeneity degree of the function
$\psi_{i}^{(0)}$ corresponding to the flat coordinate~$w$ is zero, the
homogeneity degree of the functions $\psi_{i}^{(k)}$, $k=1,\dots,N$,
corresponding to the first half of flat coordinates $u^{k}$ is $-1/2$, and
the homogeneity degree of the functions $\psi_{i}^{(N+k)}$, $k=1,\dots,N$,
corresponding to the second half of flat coordinates~$\eta^{k}$ is $1/2$. It
follows by Theorems~\ref{th-ego} and~\ref{th-const} that the
Yajima--Oikawa--Mel'nikov system has also second (local) and third
(nonlocal, with constant curvature metric) Hamiltonian structures.

\subhead{Acknowledgments} The authors are keenly grateful to
E.~V.~Ferapontov for numerous stimulating discussions, to V.~L.~Alekseev for
comments on the Whitham method and the algebro-geometric approach to the
averaging of $N$-phase solutions of the KdV and NLS equations and also to
our teacher Sergei Petrovich Novikov for interest in the paper and critical
remarks.


\begin{thebibliography}{99}
\bibitem{Aleks-95} {\normalsize V. L. Alekseev, ``On nonlocal Hamiltonian
operators of hydrodynamic type associated with the Whitham equation,'' Usp.
Mat. Nauk, \textbf{50}, No.~6, 165--166 (1995). }

\bibitem{DN-1983} {\normalsize B. A. Dubrovin and S. P. Novikov,
``Hamiltonian formalism of one-dimensional systems of the hydrodynamic type
and the Bogolyubov--Whitham averaging method,'' Dokl. Akad. Nauk SSSR,
\textbf{270}, No.~4, 781--785 (1993). }

\bibitem{Egorov} {\normalsize D. F. Egorov, Works in Differential Geometry
[in Russian], Nauka, Moscow, 1970. }

\bibitem{Zakharov} {\normalsize V. E. Zakharov, ``Benney equations and the
semiclassical approximation in the inverse problem method,'' Funkts. Anal.
Prilozhen., \textbf{14}, No.~2, 15--24 (1980). }

\bibitem{MF-90} {\normalsize O. I. Mokhov and E. V. Ferapontov, ``Nonlocal
Hamiltonian operators of hydrodynamic type that are connected with metrics
of constant curvature,'' Usp. Mat. Nauk, \textbf{45}, No.~3, 191--192
(1990). }

\bibitem{Pav-1987} {\normalsize M. V. Pavlov, ``The nonlinear Schr\"odinger
equation and the Bogolyubov--Whitham averaging method,'' Teor. Mat. Fiz.,
\textbf{71}, No.~3, 351--356 (1987). }

\bibitem{Pav-Ellips} {\normalsize M. V. Pavlov, ``Elliptic coordinates and
multi-Hamiltonian structures of hydrodynamic type systems,'' Dokl. Ross.
Akad. Nauk, \textbf{339}, No.~1, 21--23 (1994). }

\bibitem{Pav-Whitham} {\normalsize M. V. Pavlov, ``Multi-Hamiltonian
structures of the Witham equations,'' Dokl. Ross. Akad. Nauk, \textbf{338},
No.~2, 165--167 (1994). }

\bibitem{Pav-Benney} {\normalsize M. V. Pavlov, ``Local Hamiltonian
structures of Benney's system,'' Dokl. Ross. Akad. Nauk, \textbf{338},
No.~1, 33--34 (1994). }

\bibitem{PavTs} {\normalsize M. V. Pavlov and S. P. Tsarev, ``Conservation
laws for the Benney equations,'' Usp. Mat. Nauk, \textbf{46}, No.~4,
169--170 (1991). }

\bibitem{Fer-FAN} {\normalsize E. V. Ferapontov, ``Differential geometry of
nonlocal Hamiltonian operators of hydrodynamic type,'' Funkts. Anal.
Prilozhen., \textbf{25}, No.~3, 37--49 (1991). }

\bibitem{Ts-DAN} {\normalsize S. P. Tsarev, ``Poisson brackets and
one-dimensional Hamiltonian systems of hydrodynamic type,'' Dokl. Akad. Nauk
SSSR, \textbf{282}, No.~3, 534--537 (1985). }

\bibitem{Ts-Izv} {\normalsize S. P. Tsarev, ``The geometry of Hamiltonian
systems of hydrodynamic type. The generalized hodograph method,'' Izv. Akad.
Nauk SSSR, Ser. Mat., \textbf{54}, No.~5, 1048--1068 (1990). }

\bibitem{Ts-diss} {\normalsize S. P. Tsarev, Differential-Geometric
Integration Methods for Systems of Hydrodynamic Type, Ph.D. Thesis, Moscow,
1993. }

\bibitem{AP-94} {\normalsize V. L. Alekseev and M. V. Pavlov, ``Hamiltonian
structures of the Whitham equations,'' Proceedings of the Conference on NLS,
Chernogolovka, 1994. }

\bibitem{Ben} {\normalsize D. J. Benney, ``Some properties of long nonlinear
waves,'' Stud. Appl. Math., \textbf{52}, 45--50 (1973). }

\bibitem{Darboux} {\normalsize G. Darboux, Le\c{c}ons sur les syst\`{e}mes
orthogonaux et les coordonn\'{e}es curvilignes, Gautier-Villars, Paris,
1910. }

\bibitem{Dubr-96} {\normalsize B. A. Dubrovin, ``Geometry of 2D topological
field theories,'' In: Integrable Systems and Quantum Groups (Montecatini
Terme, 1993), M.~Francaviglia, S.~Greco, eds., Lecture Notes in Math.,
Vol.~1620, 1996, pp.~120--348. }

\bibitem{19} {\normalsize B. A. Dubrovin, ``Integrable systems and
classification of $2D$ topological field theories,'' In: Integrable Systems,
The J.-L. Verdier Memorial Conference, Actes du Colloque International de
Luminy (O. Babelon, P. Cartier, and Y. Kosmann-Schwarzbach, eds.),
Birkh\"auser, 1993, pp.~313--359. }

\bibitem{FP-95} {\normalsize E. V Ferapontov and M. V. Pavlov,
``Quasiclassical limit of Coupled KdV equations. Riemann invariants and
multi-Hamiltonian structure,'' Phys. D, \textbf{52}, 211--219 (1991). }

\bibitem{FL} {\normalsize M. G. Forest and J. E. Lee, ``Geometry and
modulation theory for the periodic nonlinear Schr\"odinger equation,'' In:
Oscillation Theory, Computation, and Methods of Compensated Compactness (C.
Dafermos et~al., eds.), IMA, Vol.~2, Springer-Verlag, New York, 1986,
pp.~35--69. }

\bibitem{FFM} {\normalsize H. Flaschka, M. G. Forest, and D. W. McLaughlin,
``Multiphase averaging and the inverse spectral solution of the Korteweg--de
Vries equation,'' Comm. Pure Appl. Math., \textbf{33}, 739--784 (1980). }

\bibitem{Gibbons} {\normalsize J. Gibbons, ``Collisionless Boltzmann
equations and integrable moment equations,'' Phys. D, \textbf{3}, No.~3,
503--511 (1981). }

\bibitem{Yavuz} {\normalsize H. Gumral and Y. Nutku, ``Multi-Hamiltonian
structure of equations of hydrodynamic type,'' J. Math. Phys., \textbf{31},
No.~11, 2606--2611 (1990). }

\bibitem{Pav-02} {\normalsize M. V. Pavlov, ``Integrable systems and metrics
of constant curvature,'' J. Nonlinear Math. Phys., \textbf{9}, Suppl.~1,
173--191 (2002). }

\bibitem{Oevel} {\normalsize W. Oevel, J. Sidorenko, and W. Strampp,
``Hamiltonian structures of Melnikov system and its reductions,'' Inverse
Problems, \textbf{9}, 737--747 (1993).}
\end{thebibliography}
\end{document}